\documentclass{article}

\PassOptionsToPackage{numbers, sort&compress}{natbib}

\usepackage{natbib}

\usepackage[final]{neurips_2022_ml4ps}

\usepackage[utf8]{inputenc} 
\usepackage[T1]{fontenc}    
\usepackage{hyperref}       
\usepackage{url}            
\usepackage{booktabs}       
\usepackage{amsfonts}       
\usepackage{nicefrac}       
\usepackage{microtype}      
\usepackage{algorithm}
\usepackage{algpseudocode}

\usepackage{times}
\usepackage{amsmath, nccmath, amssymb, bm}
\usepackage{amsthm}
\usepackage{geometry}
\usepackage{comment}
\usepackage{enumitem}
\usepackage{color}
\usepackage{graphicx}
\usepackage{bbold}
\usepackage{xfrac}
\usepackage{thmtools}
\usepackage{mathtools}
\usepackage{xcolor}

\newtheorem{remark}{Remark}

\DeclareMathOperator*{\argmin}{\ensuremath{\text{\rm arg\,min}}}

\DeclareMathOperator{\tr}{\ensuremath{\text{\rm tr}}}

\providecommand{\norm}[1]{\lVert#1\rVert}

\newcommand{\R}{\mathbb R}
\newcommand{\C}{\mathbb C}
\newcommand{\N}{\mathbb N}

\newcommand{\RKHS}{\mathcal{H}}
\newcommand{\Hspace}{\mathbb H}

\title{Learning dynamical systems: an example from open quantum system dynamics.}

\author{%
  Pietro Novelli\\
  Istituto Italiano di Tecnologia\\
  \texttt{pietro.novelli@iit.it} \\
}

\begin{document}

\maketitle

\begin{abstract}
Machine learning algorithms designed to learn dynamical systems from data can be used to forecast, control and interpret the observed dynamics. In this work we exemplify the use of one of such algorithms, namely Koopman operator learning, in the context of open quantum system dynamics. We will study the dynamics of a small spin chain coupled with dephasing gates and show how Koopman operator learning is an approach to efficiently learn not only the evolution of the density matrix, but also of {\em every physical observable} associated to the system. Finally, leveraging the spectral decomposition of the learned Koopman operator, we show how symmetries obeyed by the underlying dynamics can be inferred directly from data.
\end{abstract}

\section{A primer on Koopman operator learning}
The Koopman operator is a mathematical object concerned with the evolution of a dynamical system. Formally it is defined as the linear operator evolving any {\em observable}\footnote{That is, a scalar function of the state of the dynamical system.} in a pre-specified set. It owes its name to the works of Bernard Koopman on the evolution of Hamiltonian systems~\cite{Koopman1931}, but current applications of the Koopman operator embrace the whole field of dynamical systems~\cite{Budisic2012,Mauroy2020, Brunton2022}. We now provide a minimalistic introduction to the Koopman operator, referring to~\cite{Mauroy2020,Kostic2022} for a mathematically rigorous presentation. 

Given a discrete dynamical system over a state space $\mathcal{X}$, the Koopman operator $U$ is defined as
\begin{equation*}
  \left(Uf\right)(x_{t}) := f(x_{t + 1}) \quad \text{for all }t \in \N,
\end{equation*}
for every observable $f:\mathcal{X} \to \R$ in a suitable set. The Koopman operator, therefore, acts by evolving the observables of the system by one step of the dynamics. The above definition is appropriate for deterministic dynamical systems, but can be consistently extended to stochastic systems via conditional expectation operators~\cite{Meyn1993}. In this work we hinge on the formulation of Koopman operator learning 
in~\cite{Kostic2022}, in which the space of observables is restricted to a reproducing kernel Hilbert space $\RKHS$~\cite{Aronszajn1950,Steinwart2008}. Within this 
setting, an estimator $\hat{U}:\RKHS \to \RKHS$ of the Koopman operator can be {\em learned} from a single dataset of observations $(x_{i}, y_{i})_{i=1}^{n}$ consisting of sampled states $x_{i} \in \mathcal{X}$ along with their one-step-forward evolutions $y_{i} \in \mathcal{X}$. 

An obvious use of the learned operator $\hat{U}$ is forecasting both future states and {\em every observable} in $\RKHS$ (see Remark~\ref{remark:observables_forecasting}). The linearity of $\hat{U}$, however, also allows to compute its spectral decomposition, and if $(\lambda_{i}, \xi_i, \psi_i)_{i=1}^{r}$ are eigen-triplets\footnote{Namely: eigenvalues, left and right eigenfunctions.} of $\hat{U}$, the evolution of an observable $f:\mathcal{X}\to \R$ can be dissected as a sum of {\em modes}
\begin{equation}\label{eq:Koopman_mode_decomposition}
  (\hat{U}^{t}f)(x_{0}) =\sum_{i=1}^{r}\lambda_{i}^{t}\gamma_{i}^f\psi_{i}(x_{0}),
\end{equation}
where $\gamma_{i}^{f}$ is a linear functional of the observable $f$ linked to the $i$-th left eigenfunction $\xi_{i}$. The Koopman mode decomposition~\cite{Budisic2012,Arbabi2017} can be used to infer properies of the dynamical system such as constant of motions, correlation and dephasing between different observables, mixing times, normal modes, metastable states and many others. These quantities are of paramount importance in the physical sciences and machine-learning-enabled tools to compute the Koopman mode decomposition unlock the possibility to study complex physical systems too hard to be approached with analytical theories alone.
 
\begin{remark}[Forecasting the observables]\label{remark:observables_forecasting} Suppose we have observed $T$ states $(x_{t})_{t \leq T}$ of a (possibly noisy) dynamical system and we are interested in learning the evolution of its observables $f:\mathcal{X} \to \R$. A straightforward approach is to first learn an map $\hat{S}$ between the states such that $x_{t + 1} \approx \hat{S}(x_{t})$ and then predict any observable by simply composing it with the learned map $\hat{S}$, i.e. $f(x_{t+1}) \approx f(\hat{S}(x_{t}))$. Here, $\hat{S}:\mathcal{X}\to\mathcal{X}$ can be anything: a neural network, a vector-valued kernel etc. The framework developed in~\cite{Kostic2022} for the Koopman operator learning, on the contrary, actively capitalize on the structure of the space of observables $\RKHS$ to {\em directly} provide the best forecast as $f(x_{t+1}) \approx (\hat{U}f)(x_{t})$ for {\em any} observable $f \in \RKHS$, without the middle step of learning the dynamical map $\hat{S}$. The two approaches coincide if the underlying dynamics is deterministic, whereas for stochastic dynamical systems the first approach is sub-optimal (i.e. leading to larger errors).
\end{remark}

\section{Quantum dynamics and the Lindblad equation}
In the context of open quantum system dynamics, the state of a quantum mechanical system is described by a {\em density matrix} (or operator) $\rho: \Hspace \to \Hspace$, where $\Hspace$ is a Hilbert space characterizing the quantum system~\cite[Chapter 2]{Nielsen2012}. Physical requirements prescribe that the density matrix is a positive definite, hermitian operator satisfying $\tr(\rho) = 1$. Any physical observable $A$ is an hermitian operator acting on $\Hspace$, and its quantum mechanical expectation value for a given state $\rho$ is $\langle A \rangle_{\rho} := \tr (A\rho)$. Noticing that $\tr(A\rho)$ is the scalar product between $A$ and $\rho$ in the Hilbert-Schmidt sense, we conclude that the observables of physical interest are {\em linear} functionals of $\rho \in {\text{HS}}(\Hspace)$, where ${\text{HS}}(\Hspace)$ is the space of Hilbert-Schmidt operators on $\Hspace$.

For quantum systems with discrete energy levels, the Hilbert space $\Hspace$ is isomorphic to $\C^{d}$ for some (possibly large) $d$. In the case of $N$ two-level systems such as spins, one has $\Hspace \sim \C^{2^{N}}$.

If the physical system is closed, i.e. isolated from any interaction with the environment, its evolution is unitary and governed by the Schr\"odinger equation. Conversely, when the system is open its equation of motion is also affected by the environment. Under proper assumptions and approximations\footnote{Namely: separability, Markovianity, Born and secular approximations.}, the dynamics of an open quantum system is described by the Lindblad equation~\cite[Chapter 3]{Lindblad1976,Wiseman2009}
\begin{equation*}
  \frac{d}{dt}\rho(t) = -\frac{i}{\hbar}\left[H,\rho(t)\right] + \sum_{k}\left[ L_{k}\rho(t) L_{k}^{\dag} - \frac{\rho(t) L_{k}^{\dag}L_{k} + L_{k}^{\dag}L_{k}\rho(t)}{2} \right],
\end{equation*}
where $H$ is the Hamiltonian of the system and $L_{k}:\Hspace \to \Hspace$ are {\em collapse operators} describing disspiative processes. The Lindblad equation is a linear ordinary differential equation for the density matrix $\rho$. Its form guarantees that for all $t$, $\rho(t)$ is a density matrix (i.e. satisfy the aforementioned requirements of positivity, hermiticity and unit trace). 
\begin{remark}[Relation with Heisenberg and Schr\"odinger pictures]
It is well known that the evolution of quantum systems can be described by two equivalent {\em pictures}. In the Schr\"odinger picture, which we employed above, the states evolve in time, while the operators are fixed. In the Heisenberg picture the opposite is true: states are fixed and operators evolve. Clearly, because of theirs equivalence, the choice of the picture do not affect the evolution of the expected value $\langle A \rangle_{\rho}(t) = \tr (A^{\rm{H}}(t)\rho) = \tr (A\rho^{\rm{S}}(t))$. The Koopman operator acts directly on the expected value $\langle A \rangle_{\rho}(t)$ and is therefore well defined in both pictures. Koopman operator learning, however, is based on a dataset of {\em evolved states} and is arguably more natural to adopt the Schr\"odinger picture to study its properties.
\end{remark}
\section{Koopman operator learning on a spin chain}
We now illustrate Koopman operator learning on a simple spin chain model by generating the training dataset from a simulated Lindblad evolution. This synthetic setting is mostly useful as a benchmark, as we can compare the {\em learned} dynamical system with the ground truth. A more sensible use of Koopman operator learning that we envision, however, is to make use of experimental measurements on complex quantum systems (e.g. quantum computers) to distill an accurate and interpretable model of theirs dynamical evolution.

We consider a spin chain with $N$ spins described by the Hamiltonian
\begin{equation}\label{eq:spin_chain_hamiltonian}
  H := -\frac{1}{2}\sum_{i=1}^{N-1} J_{\parallel}\left(\sigma_{i}^{x}\sigma_{i+1}^x + \sigma_{i}^{y}\sigma_{i+1}^y\right) +J_{\perp}\sigma_{i}^{z}\sigma_{i+1}^z,
\end{equation}
where $\sigma_{i}^{x,y,z}$ are Pauli matrices corresponding to the $i$-th spin. The model~\eqref{eq:spin_chain_hamiltonian} is also known as quantum Heisenberg model, and under proper assumptions on the coupling constants $J_\parallel, J_\perp$ can be exactly solved using the Bethe ansatz~\cite{Karbach1997,Karbach1998}. 

We let the spin chain~\eqref{eq:spin_chain_hamiltonian} interact with the environment via dephasing channels~\cite{Marquardt2008}. To this end, we choose the collapse operators of the Lindblad equation as $N$ pure dephasings $L_{i} = \sigma^{z}_{i}\sqrt{\Gamma/2}$ with rate $\sqrt{\Gamma/2}$. As explained in~\cite{Marquardt2008}, a pure dephasing $\sigma^{z}\sqrt{\Gamma/2}$ for a two level system is responsible for the decay of the off-diagonal elements of the density matrix at a rate $\propto e^{-\Gamma t}$.

\paragraph*{Experimental details.} The data are generated by simulating a chain of $N=5$ spins with the QuTiP framework~\cite{Johansson2012}. We construct a dataset of quantum states by saving the density matrix of the simulated system $\rho(t)$ at times $t = k\Delta t$ with $k = 1, \ldots, 200$ and $\Delta t := 0.5$. The first half of the data, corresponding to the initial 100 timesteps, is used to train the Koopman operator estimator $\hat{U}$, while the second half is used for testing. The initial state $\rho(0)$ for the integration of the Lindblad equation is chosen as the pure state $\vert\downarrow \uparrow\uparrow\uparrow\uparrow\rangle$, corresponding to the first spin down and the rest up. The coupling constants and dephasing rate are set to $J_{\parallel} := 0.1\pi$, $J_{\perp} := 0.2\pi$, and $\Gamma = 0.01$, respectively.

The estimator $\hat{U}$ of the Koopman operator is computed using the {\em reduced rank regression} algorithm presented in~\cite{Kostic2022} with rank $r=19$, Tikhonov regularization $\lambda = 10^{-6}$ and a linear kernel. The {\em reduced rank regression} algorithm with linear kernel on the dataset $(x_{i}, y_{i})_{i=1}^{n}$ returns the minimizer
\begin{equation*}
  \hat{U} := \argmin_{U \in {\text{HS}}_{r}(\mathcal{X})} \frac{1}{n}\sum_{i=1}^{n} \norm{y_{i} - U^* x_{i}}^{2} + \lambda \norm{U}_{{\text{HS}}}^{2},
\end{equation*}
where ${\text{HS}}_{r}(\mathcal{X})$ is the set of Hilbert-Schmidt operators on $\mathcal{X}$ with rank smaller or equal than $r$ and $U^*$ is the hermitian conjugate of $U$.

The code to reproduce the experiments is available at the link \href{https://github.com/CSML-IIT-UCL/kooplearn}{https://github.com/CSML-IIT-UCL/kooplearn}. The experiments have been conducted on a workstation equipped with an Intel(R) Core\textsuperscript{TM} i9-9900X CPU @ 3.50GHz, 48GB of RAM and a NVIDIA GeForce RTX 2080 Ti GPU.

\subsection{Forecasting the physical observables}
We now compare the dynamics of selected physical observables as forecasted by the Koopman operator estimator $\hat{U}$ against the ground truth obtained by direct integration of the Lindblad equation. We focus on the spin polarization and on the spin current. The spin polarization (along $z$) of the $i$-th spin is the observable associated with the spin operator $S_{i}^{z} := \hbar \sigma_{i}^{z}/2$. A non-vanishing spin polarization $\tr(S_{i}^{z}\rho)$ across many sites signals the onset of a magnetic order for the state $\rho$. The Heisenberg model~\eqref{eq:spin_chain_hamiltonian}, indeed, is a rudimentary model of a magnetic material and depending on the coupling constants $J_{\parallel}$ and $J_{\perp}$ might exhibit either ferromagnetic or antiferromagnetic order~~\cite{Karbach1997,Karbach1998}. 

The spin current at site $i$ can be derived from the continuity equation as detailed in~\cite{Schutz2004} and reduces to the operator
\begin{equation*}
  j_{i}^{z} := \frac{iJ_{\parallel}}{4}\left[(\sigma^{+}_{i}\sigma^{-}_{i + 1} - \sigma^{-}_{i}\sigma^{+}_{i + 1}) - (\sigma^{+}_{i-1}\sigma^{-}_{i} - \sigma^{-}_{i - 1}\sigma^{+}_{i}) \right],
\end{equation*}
where $\sigma_{i}^{\pm} := \sigma_{i}^{x} \pm i\sigma_{i}^{y}$ are the usual ladder operators for the $i$-th spin. The first term in round parenthesis in the definition of $j_{i}^{z}$ accounts for {\em outgoing} currents while the second term for {\em incoming} currents at site $i$. Manipulation and control of spin currents is of great interest in the field of {\em spintronics}, dealing with the use electrons' spin degrees of freedom for information processing. In spintronic devices, data storage and transmission can be achieved at higher efficiency compared to standard electronic devices~\cite{Hirohata2020}.

In Figure~\ref{fig:observables_forecast} we show the forecast of, respectively, the average spin polarization and the average spin current of selected sites along the chain. In both panels, the left portion of the plot corresponds to the data used to learn the Koopman operator estimator $\hat{U}$, while in the right portion of the plot are compared the forecasted dynamics on unseen data against the true dynamics. The forecast is accurate both for spin polarization and current at short times (i.e. for $50 \leq t \lessapprox 70$). At later times, however, the forecasted spin polarization diverge from the ground truth, while the forecasted spin current stays remarkably close to the true spin current. 
{\small
\begin{figure}[t!]
\begin{center}
\includegraphics[width=0.9\textwidth]{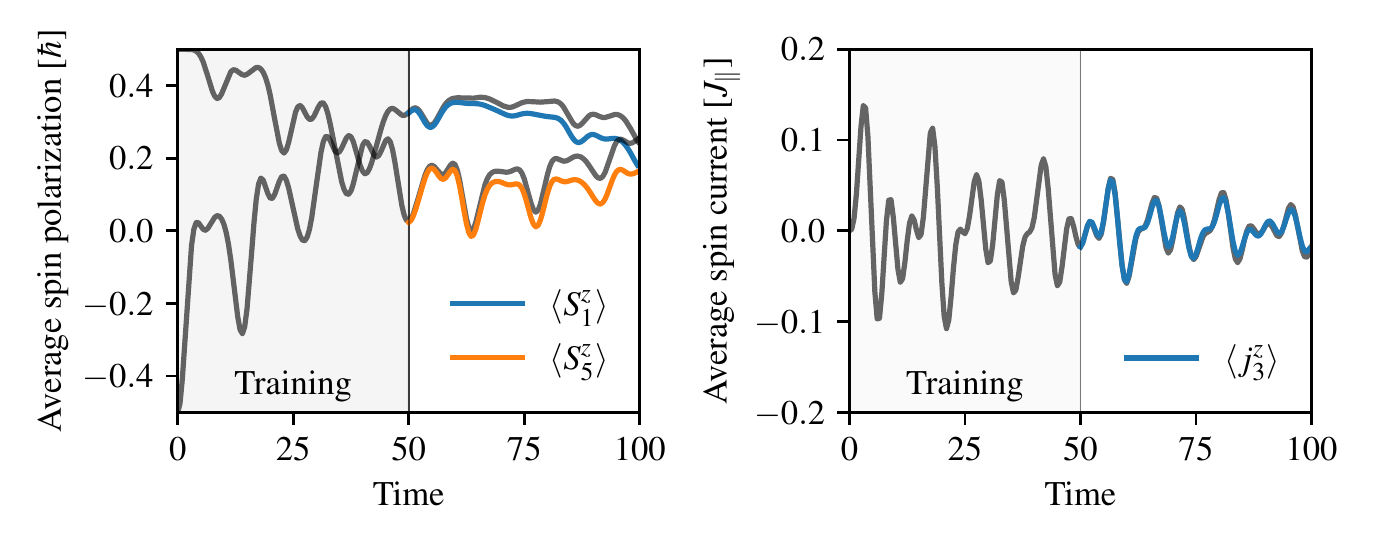}
\end{center}
    \caption{Left panel: forecast of the average spin polarization of the first (1) and last (5) spin of the chain in units of $\hbar$. Right panel: forecast of the average spin current of the central (3) spin of the chain in units of $J_{\parallel}$. In both panels the gray traces correspond to the ground truth and the first half of the data were used to train the Koopman operator estimator. }\label{fig:observables_forecast}
\end{figure}
}

\subsection{Decay rates, normal modes, constants of motion and symmetries}
We now highlight how the Koopman mode decomposition~\eqref{eq:Koopman_mode_decomposition} might be used to infer quantities of physical interest such as decay rates, frequency of the normal modes and constants of motion of the dynamics. 

Decay rates and frequency of the normal modes are easily obtained noticing that the time-dependence in~\eqref{eq:Koopman_mode_decomposition} is expressed only by the eigenvalues $\lambda_{i}$ of $\hat{U}$. The theory of dynamical systems dictates that $|\lambda_{i}| \leq 1$ for stable systems. The powered eigenvalues $\lambda_{i}^{t}$ appearing in~\eqref{eq:Koopman_mode_decomposition} can be conveniently expressed in polar form, yielding $\lambda_{i}^{t} := e^{-t/\tau_{i}}e^{i2\pi\omega_{i} t}$, with $\tau_{i} = -t/\log(|\lambda_{i}|)$ and $\omega_{i} = \arg(\lambda_{i})/(2\pi t)$. 

Computing the decay rates of the learned $\hat{U}$ we get that 16 out of 19 eigenvalues have decay rates in the interval $[0.018, 0.021]$, which is quite close to the true rate $\Gamma = 0.01$ selected for the pure dephasing gates. The remaining three decay rates are, respectively $0.014, 0.026$ and $0.003$, the last one of which corresponds to the steady-state solution. Similarly, the computed frequencies $\omega_{i}$ lie in the range $[0.05, 0.2]$, of the same order of the (properly normalized) coupling constants $J_{\parallel}/2\pi = 0.05$ and $J^{z}/2\pi = 0.1$.

The last point we would like to comment concerns the mode corresponding to the steady-state of the system, i.e. the one with eigenvalue one. This mode, in practice, is characterized by an {\em estimated} Koopman eigenvalue very close to $1$ and in our calculation is given by $\lambda_{1} = 0.9985$. By construction, the right eigenfunction associated to the steady-state mode is a constant of motion. Indeed, $\psi_{1}(\rho_{t}) = U{t}\psi_{1}(\rho_0) = \lambda_{1}^{t}\psi_{1}(\rho_{0}) = \psi_{1}(\rho_{0})$ for all $t$. 

In the present case we are dealing with {\em linear} observables\footnote{That is, linear functionals of the density matrix.}, and every eigenfunction $\psi_{i}(\rho)$ can be expressed, by construction, as $\psi_{i}(\rho) = \tr(\Psi_{i}\rho)$ for some operator $\Psi_{i}$. For the steady-state mode, the operator $\Psi_{1}$ is a symmetry transformation for which the dynamics is invariant. We verified that the estimated $\Psi_{1}$ commutes with the total spin operator along $z$, as expected for the pure dephasing channel~\cite{Marquardt2008}.

{\em Acknowledgments}: I wish to tank P. Erdman for useful discussions and for his comments on the manuscript.
\bibliographystyle{apalike}
\bibliography{bibliography}
\end{document}